\documentclass[12pt]{iopart}
\usepackage{graphicx}
\usepackage{iopams}

\newcommand{\bk}{{\bf k}}

\newcommand{\cZ}{{\mathcal Z}}

\begin{document}

\title[Preprint submitted to J. Stat. Mech.]{Stochastic simulations for the time evolution of systems which
obey generalized statistics: Fractional exclusion statistics
and Gentile's statistics}

\author{G. A. Nemnes}

\address{
University of Bucharest, Faculty of Physics,
``Materials and Devices for Electronics and Optoelectronics'' Research Center,
P.O. Box MG-11, 077125 Magurele-Ilfov, Romania}
\ead{nemnes@solid.fizica.unibuc.ro}

\author{D. V. Anghel}
\address{
``Horia Hulubei''
National Institute for Physics and Nuclear Engineering (IFIN-HH),
077126 Magurele-Ilfov, Romania}
\ead{dragos@theory.nipne.ro}

\begin{abstract}
We present a stochastic method for the simulation of the time evolution 
in systems which obey generalized statistics, namely
fractional exclusion statistics and Gentile's statistics.
The transition rates are derived in the framework of canonical ensembles.
This approach introduces a tool for describing 
interacting fermionic and bosonic systems in non-equilibrium as ideal FES
systems, in a computationally efficient manner.
The two types of statistics are analyzed comparatively, indicating 
their intrinsic thermodynamic differences and revealing key aspects
related to the species size. 
\end{abstract}

\maketitle

\section{Introduction\label{intro}}

Two of the best known generalizations of the Bose and Fermi
statistics are the Gentile's statistics (GS)
\cite{NuovoCim.17.495.1940.Gentile,JMathPhys.11.2691.1970.Katsura,RevBrasFis.6.471.1976.Ponczek}
and the fractional exclusion statistics (FES)
\cite{PhysRevLett.67.937.1991.Haldane}. 
A method for the stochastic simulation of the time evolution of Bose and 
Fermi gases was proposed by Guastella \textit{et al.} in Ref. 
\cite{JStatMech.2009.P02021.2009.Guastella}.
In this paper we present a method for the simulation of 
the more general GS and FES systems,
which have as limit cases the Fermi and Bose statistics.
As we shall see below, the GS and the FES are apparently 
closely related. Nevertheless they give totally different
thermodynamic results. 

The GS describes systems of non-interacting particles in which each
single particle state can be occupied by a maximum of $\xi$ particles;
the case $\xi=1$ corresponds to the Fermi statistics and the case
$\xi=\infty$ corresponds to the Bose statistics. 

The definition of FES is somewhat more complicated. A FES system is
formed in general of several subsystems, called \textit{species}. Each
species consists of a finite number of particles in a finite
dimensional space, spanned by a set of single-particle quantum
numbers. They are defined by coarse-graining the set of
single-particle quantum numbers for each type of
particles in the system. 
Each species has a certain number of available
single-particle states which may depend, in principle, on the number
of particles in any of the species of the system. 
This dependence is known as the
``statistical interaction'' between the FES particles and limits the
number of particles that can be accommodated in each species,
similarly as in the GS.

Let us number the species by $i=0,1,\ldots$ and denote by
$N_i$ and $G_i$ the number of particles and the number of available
single-particle states of  species $i$, respectively. The
statistical interaction is expressed by the change in the number of available
single-particle states at the change of the particle numbers. For
example a change $\delta N_i$ of $N_i$ will change $G_j$ by
$-\alpha_{ji}\delta N_i$, for any $j$. The FES parameters,
$\alpha_{ij}$, depend on the way the system is divided into species 
-- different divisions lead to different $\alpha$'s 
\cite{EPL.87.60009.2009.Anghel}.
For non-interacting bosons and fermions, $\alpha_{ij}=0$ and
$\alpha_{ij}=\delta_{ij}$, respectively, for any $i$ and $j$.

Now the connection between GS and FES is obvious. If we would have
two similar types of systems, one described by GS, of parameter
$\xi$, and the other one described by FES, of parameters
$\alpha_{ij}=\delta_{ij}\alpha\equiv\delta_{ij}/\xi$, we may
coarse-grain the sets of quantum numbers that describe  these systems
and obtain the ``species'' $i=0,1,\ldots$, each with
$N_i$ particles and $G_i$ available states. If we denote by
$G^{(0)}_i$ the number of
available states when $N_i=0$, for any $i$, then the maximum number of particles
that we can put into the species $i$ in both gases is the same, namely $\xi
G^{(0)}_i=G^{(0)}_i/\alpha$. 
Based on this single observation we could say that GS is pure FES with
species formed of single-particle states.

Nevertheless, the thermodynamics of FES and GS systems are
different. In general, FES is applied to species of
large numbers of particles and states and cannot be applied to
single states, like GS. If one would attempt to make the FES species
  smaller and smaller, not only that the thermodynamic results deviate
  from those obtained with large species, as we shall show below, but
  the very definition of FES becomes unclear. On the other hand, if we
  apply both, FES and GS to particle species--say we have the species
  $\{N_i,G_i\}_{i=0,1,\ldots}$--the number of microconfigurations in
  which the particles can arrange themselves is not the same in FES
  \cite{PhysRevLett.67.937.1991.Haldane,PhysRevLett.73.922.1994.Wu,PhysRevLett.73.2150.1994.Isakov,JPhysA.40.F1013.2007.Anghel,EPL.90.10006.2010.Anghel}
  and GS
  \cite{NuovoCim.17.495.1940.Gentile,JMathPhys.11.2691.1970.Katsura,RevBrasFis.6.471.1976.Ponczek}
  systems. Therefore the thermodynamic results must be different.

The difference between the FES and the GS is very clearly evidenced in systems
of constant density of states (DOS). It is well known that Bose,
Fermi and, in general, FES systems of constant DOS are
thermodynamically equivalent, i.e. the specific heat and entropy are independent of the
statistics \cite{ProcCambrPhilos42.272.1946.Auluc,PhysRev.135.A1515.1964.May,PhysRevE.55.1518.1997.Lee,PhysA304.421.2002.Lee,JPA35.7255.2002.Anghel,JPA39.4787.2006.Anghel}. 
The microscopic reason of this equivalence lies in the fact that one 
can realize a one-to-one mapping between microconfigurations of particles 
of the same excitation energies across all statistics -- by the 
``excitation energy'' of a microconfiguration we mean the difference between the 
energy of the microconfiguration and the lowest energy of the system 
\cite{JPA35.7255.2002.Anghel}.  
As we shall see in the following, the GS does not fall into
this equivalence class. For a GS of constant DOS, the specific heat,
and therefore also the entropy, depend on the parameter $\xi$. 


As stated above, we propose in this paper a method to stochastically 
simulate the time evolution of GS and FES systems. 
The paper is organized as follows. For the clarity of the exposition, we
start in Section \ref{BFgas} by briefly reviewing the method of
Guastella \textit{et al.}. Then we extend the method to the FES and
GS gases in Section 3. Finally, in Section 4
the derived transition probabilities associated with the Markov chain 
that we introduce
are used in simulations of FES and GS systems, the results are tested 
against corresponding analytical data 
and particularities of the two statistics are commented upon.

\section{Simulation of Bose and Fermi ideal gases \label{BFgas}}

Let us consider a system of ideal bosons or fermions in contact with a heat
bath, at temperature $T$. The single particle states of the system, 
denoted by $|i\rangle$, have the energies
$\epsilon_i$ ($i=0,1,\ldots$); the energy levels may be
degenerated. 

The dynamics of the system is described in terms of a
time dependent Markov chain in the microconfiguration space.
Each microconfiguration is a set $\{n_i\}_{i=0,1,\ldots}$, 
where $n_i$ is the particle population of the state $|i\rangle$.
For fluency we shall call the microconfigurations simply {\it states},
which must not be confused with the single-particle states.  
The system evolves in time from one state to another {\it neighboring}
state by random jumps of particles between the single-particle
states. The transition rates between two states will be denoted by
$T_{\{n'_i\},\{n_i\}}$, where the notation 
convention is that the first subscript refers to the final state and 
the second subscript refers to the initial state.
Assuming that in the dynamics of the system 
the particles jump between single particle states
one at a time, we shall consider that two states are neighbors if they
differ by the exchange of only  
one particle between two single particle states. Therefore, we shall use the simpler 
notation $T_{ji}(n_j,n_i)$ to refer to the transition rate of a particle from the state 
$|i\rangle$, of $n_i$ particles, to the state $|j\rangle$, of $n_j$ particles.

If the particles are bosons, the Fermi golden rule
imposes in general that $T_{ji}(n_j,n_i)$ is proportional to
$(n_j+1)n_i$.
For fermions, the Pauli exclusion principle gives 
a transition rate $T_{ji}\propto (1-n_j)n_i\equiv n_i\delta_{n_j,0}$, 
since each state can accommodate no more than one particle.
As the system is in contact with a thermostat, the
\textit{relative} transition rate, $T_{ji}(n_j,n_i)/T_{ij}(n_i,n_j)$, must also 
be proportional to a weight factor, 
$e^{-\beta(\epsilon_j-\epsilon_i)}$, where $\beta\equiv1/(k_BT)$. 

After setting-up the initial microconfiguration, the simulation 
proceeds by allowing the particles to make random jumps between 
the single-particle states, with probabilities proportional to the transition 
rates. 
In what follows we shall denote by $p_{\{n_i\}}$ the probability
  to find the system in the microconfiguration $\{n_i\}$. Then the
mean occupation of a single particle state is 
\begin{equation}
\langle n_j \rangle = \sum_{\{n_i\}} p_{\{n_i\}} n_j.  \label{averagenj}
\end{equation}

After the system is simulated long enough, the equilibrium is
reached and the ensemble averages of the thermodynamic quantities do
not change in time anymore. Once this condition is reached, the equilibrium probabilities, 
$p^{\rm Eq}_{\{n_i\}}$, satisfy the detailed balance equation (DBE), 
\begin{equation}
p^{\mbox{\scriptsize Eq}}_{\{n_k\}} T_{\{n'_k\},\{n_k\}} 
= p^{\mbox{\scriptsize Eq}}_{\{n'_k\}} T_{\{n_k\},\{n'_k\}}. \label{DBE1}
\end{equation}
for any neighboring states, $\{n_k\} \equiv ( \ldots, n_i, \ldots, n_j, \ldots)$ and 
$\{n'_k\} \equiv ( \ldots, n'_i, \ldots, n'_j, \ldots) \equiv 
( \ldots, n_i-1, \ldots, n_j+1, \ldots)$. 
On the other hand, since the system is in contact with a heat 
bath, the equilibrium probabilities should be the canonical 
probabilities, 
\begin{equation}
  p^{\rm Eq}_{\{n_i\}} = e^{-\beta[\sum_{i}n_i\epsilon_i]}/Z(T,N), \label{pcan}
\end{equation}
where $Z(T,N)$ is the canonical partition function. 

The DBE can be easily checked. If we plug the expression (\ref{pcan}) into 
(\ref{DBE1}), with $T_{\{n'_k\},\{n_k\}}\equiv T_{ji}(n_j,n_i)\propto (1\pm n_j)n_i$, 
we obtain an identity. 
Therefore the procedure described above produces the canonical 
probabilities when equilibrium is reached.


\section{Simulation of systems which obey intermediate statistics}
\label{formalism}

\subsection{Fractional exclusion statistics\label{FESsection}}

The calculation of the transition rates in FES systems is not as
straightforward as in the Bose and Fermi gases. A FES system is
essentially an interacting particle system, described 
by a (countable) set of ``quasiparticle'' quantum numbers, $\bk_i$,
where $i=0,1,\ldots$
\cite{PhysRevLett.67.937.1991.Haldane,PhysRevLett.73.922.1994.Wu,PhysRevLett.73.2150.1994.Isakov,PhysLettA.372.5745.2008.Anghel,RJP.54.281.2009.Anghel,EPL.90.10006.2010.Anghel}. 
For example 1D integrable quantum gases of interacting bosons and
fermions 
\cite{NewDevIntSys.1995.Bernard,PhysRevB.56.4422.1997.Sutherland,EPL.90.10006.2010.Anghel}
and systems described by the Fermi liquid model  
\cite{PhysRevLett.73.3331.1994.Murthy,PhysRevLett.74.3912.1995.Sen,JPA35.7255.2002.Anghel,RomRepPhys59.235.2007.Anghel,PhysLettA.372.5745.2008.Anghel,EPL.90.10006.2010.Anghel}
are typical FES systems. For the 1D integrable quantum gases the
quantum numbers are the asymptotic momenta, $k_i$, whereas for the Fermi
liquid systems the ``quasiparticle'' energies, $\tilde\epsilon_i$, are
used as quantum numbers. The population of the ``quasiparticle''
states will be denoted by $n_i$ or $n(k_i)$. The particles are
interacting and therefore the values
of the quantum numbers may depend (in principle) on the whole set
$\{n_i\}_{i=0,1,\ldots}$, which defines also the quantum state of the system.

To show how the dependence of $\bk_i$ on 
$\{n_j\}_{j=0,1,\ldots}$ leads to FES (see also \cite{RJP.54.281.2009.Anghel} for more 
details), let us assume that the allowed values
of $\bk_i$, $\{\bk_i\}_{i=0,1,\ldots}$, form a
set of points in a $d$-dimensional ($d$D) vector
space. To build a
statistical description of the system, we have to calculate the number
of configurations in which the particles can occupy different one-particle states at
constant total energy $E$ and particle number $N$. For this, we divide the
$d$D space of $\bk$ vectors into elementary, fixed volumes, $V_i$, each
containing $G_i$ states and $N_i$ particles -- these are our species and
should be made no confusion from using the same subscripts, $i,j$, etc. for
counting both, the states, $|\bk_i\rangle$, and the species,
$V_i,G_i,N_i$. If the particles in the system are bosons, $n_i$ may
take any positive integer value and the number of microconfigurations
we have in the system for this division into species is 
%
\numparts  \label{N_microC}
\begin{equation}
W_B(\{G_i,N_i\}) = \prod_i \frac{(G_i+N_i-1)!}{N_i!(G_i-1)!} . 
\label{Bose_microC}
\end{equation}
If the particles are fermions, $n_i=0$ or 1 and the total number of
microconfigurations is 
\begin{equation}
W_F(\{G_i,N_i\}) = \prod_i \frac{G_i!}{N_i!(G_i-N_i)!} . 
\label{Fermi_microC}
\end{equation}
\endnumparts
%

Since in our system the values $\{\bk_i\}$ change with the particle
populations, $\{n_i\}$, while the elementary
volumes, $V_i$, are 
fixed in the $\bk$ space, this leads to a dependence of the $G_i$'s on
the set $\{N_j\}_{j=0,1,\ldots}$ \cite{JPhysA.40.F1013.2007.Anghel,PhysLettA.372.5745.2008.Anghel,RJP.54.281.2009.Anghel,EPL.90.10006.2010.Anghel}. 

To quantify this statement, let us assume that a small change,
$\delta N_i$, of $N_i$ produces a linear change,
\begin{equation}
\label{deltaGj}
\delta G_j=-\alpha_{ji}\delta N_i,  
\end{equation}
of the number of states, $G_j$,
for any $i$ and $j$  (not all the $\alpha$'s have to be different from
zero). The
proportionality constants, $\alpha_{ij}$, are the FES parameters 
defined in the Introduction.

Using Eqs. (\ref{N_microC}) and (\ref{deltaGj}) we can calculate the grandcanonical partition
function, 

\begin{equation}
\cZ^{\rm FES}_p (\{G_i,N_i\}) = 
\sum_{\{N_i\}}e^{\beta\sum_iN_i(\mu-\epsilon_i)} W_p (\{G_i,N_i\})
\label{cZ1}
\end{equation}
where we assumed that the total energy of the system is
$E(\{N_i\})\equiv\sum_iN_i\epsilon_i$. This is equivalent to ascribing
the same energy--the average energy--to all the particles in a
species. In Eq. (\ref{cZ1}) $p$
stands for $B$ or $F$, depending on
whether we have bosons or fermions in the system.
Using Eq. (\ref{cZ1}) we can calculate the equilibrium properties of
the system by maximizing $\cZ^{\rm FES}$ with respect to the set $\{N_i\}$
\cite{PhysRevLett.73.922.1994.Wu,EPL.90.10006.2010.Anghel}. 
In this way we obtain for bosons and fermions the systems of equations 
%
\numparts \label{eqplesdistr}
\begin{equation}
\frac{1+\tilde n_i}{\tilde n_i}\prod_j(1+\tilde n_j)^{-\alpha_{ji}} = e^{\beta(\epsilon_i-\mu)}
\label{nBose}
\end{equation}
\begin{equation}
\frac{1-\tilde n_i}{\tilde n_i}\prod_j(1-\tilde n_j)^{\alpha_{ji}} = e^{\beta(\epsilon_i-\mu)} ,
\label{nFermi}
\end{equation}
\endnumparts
%
respectively, where $\tilde n_i\equiv N_i/G_i$. 
One should note that in both Eqs. (\ref{nBose}) and (\ref{nFermi})
$\alpha_{ji}=0$ corresponds to non-interacting bosonic and fermionic systems,
respectively.

To calculate the transition rates, we have to consider separately the systems of 
bosons and the systems of fermions; we shall be interested in transitions 
in which a particle jumps from one species to another, therefore changing the 
particle distribution over the species. 

Let's consider the jump of one particle from the species $i$ into the species $j$ 
and vice-versa, in a system of bosons. For the jump $i\to j$, we start from a 
configuration of $N_i$ particles and $G_i$ available states in species $i$ and arrive into a 
configuration of $N_j$ particles and $G_j$ states in the species $j$. The reversed process, 
$j\to i$, is exactly like $i\to j$, but in reverse order, i.e. we start from a 
configuration of $N_j$ particles and $G_j$ states in the species $j$ and arrive 
in a configuration of $N_i$ particles and $G_i$ states in the species $i$. 

Let's now focus on the process $i\to j$ and assume that the $N_i$ particles 
are distributed as $n_{i_0},n_{i_1},\ldots$ onto the states with the quantum numbers 
$\bk_{i_0},\bk_{i_1},\ldots$. 
Analogously, the $N_j-1$ particles that exist in the species $j$ before the jump, are
distributed as $n_{j_0},n_{j_1},\ldots$ particles on the states 
$\bk_{j_0},\bk_{j_1},\ldots$. Keeping the notations 
from Section \ref{BFgas}, we have the transition rates 
$T_{j_m i_n}(n_{j_m},n_{i_n})\propto(n_{j_m}+1)n_{i_n}$. Therefore the transition rate 
of a particle--any particle--from the species $i$ to the state $\bk_{j_m}$ of 
species $j$ is $T_{j_m i}(n_{j_m},N_i)\propto(n_{j_m}+1)N_i$. Moreover, the transition 
of a particle from the species $i$ onto any state of the species $j$ can be found 
by summing-up over all the $G_j$ states of the species $j$, leading to 
%
$T_{B,ji}(N_j,N_i) \propto (G_j+N_j-1)N_i$. 
%

Further, to obtain the total transition rate from the species $i$ to the species $j$, 
we have to multiply the above result with the number of microconfigurations, 
$W_{B,ij}=(G_j+N_j-2)!/[(N_j-1)!(G_j-1)!] \times (G_i+N_i-1)!/[N_i!(G_i-1)!]$. 
In this way we finally obtain, up to a common multiplicative factor and the 
canonical weight, $e^{\beta(\epsilon_i-\epsilon_j)}$, 
\begin{equation}
T_{B,ji}(N_j-1,N_i) = \frac{(G_j+N_j-1)!}{(N_j-1)!(G_j-1)!}
\frac{(G_i+N_i-1)!}{(N_i-1)!(G_i-1)!}. 
\label{TBji}
\end{equation}

The DBE then reads, 
\begin{equation}
\fl
e^{-\beta[N_i\epsilon_i+(N_j-1)\epsilon_j]}T_{B,ji}(N_j-1,N_i) e^{\beta(\epsilon_i-\epsilon_j)}
=e^{-\beta[(N_i-1)\epsilon_i+N_j\epsilon_j]}T_{B,ij}(N_i-1,N_j) 
\label{BDEFESB}
\end{equation}
which, by using Eq. (\ref{TBji}), it can be immediately checked to be an identity.

A similar argument can be used for transitions in systems of fermions,
with the difference that  the particles cannot arrive on a quantum
state already occupied. This leads to a transition rate 
%
$T_{F,ji}(N_j,N_i) \propto (G_j-N_j)N_i$. 
%
Multiplying again by the number of microconfigurations, we obtain 
\begin{equation}
T_{F,ji}(N_j-1,N_i) = \frac{G_j!}{(N_j-1)!(G_j-N_j)!}
\frac{G_i!}{(N_i-1)!(G_i-N_i)!}. 
\label{TFji}
\end{equation}

Like in the Bose systems, the transition rates (\ref{TFji}) satisfy identically the 
DBE.

\subsection{Gentile's statistics\label{GSsection}}

In GS each single-particle state can be occupied by up to $\xi$ particles. This can be 
described in a second quantized theory by introducing the creation and annihilation operators 
$a_\bk^\dagger$ and $a_\bk$, respectively, of the form 
\cite{JMathPhys.11.2691.1970.Katsura}
\begin{eqnarray}
\fl
a_\bk^\dagger = \left(\begin{array}{ccccccc}
0 & 0 & & & & & \\
1 & 0 & & & & & \\
0 & \sqrt{2} & & & & & \\
  & & \cdot & & & & \\
  & & & \cdot & & & \\
  & & & & \cdot & & \\
  & & & & & \sqrt{\xi} & 0
\end{array}\right)_\bk ,\ 
a_\bk = \left(\begin{array}{ccccccc}
0 & 1 & 0 & & & & \\
0 & 0 & \sqrt{2} & & & & \\
 & & & \cdot & & & \\
 & & & & \cdot & & \\
 & & & & & \cdot & \\
 & & & & & & \sqrt{\xi} \\
 & & & & & & 0
\end{array}\right)_\bk .
\label{defaaGS}
\end{eqnarray}
The creation and annihilation operators are assumed to commute for different $\bk$'s. 
If we denote by $|0\rangle$ the vacuum state, then a many-body state of 
$N=n_{\bk_0}+n_{\bk_1}+\ldots$ particles is 
$|n_{\bk_0},n_{\bk_1},\ldots\rangle=(a_{\bk_0}^\dagger)^{n_{\bk_0}}(a_{\bk_1}^\dagger)^{n_{\bk_1}}\ldots|0\rangle$. 

As the system is connected to a heat bath, the 
eigenstates of the free GS particle system are perturbed 
by external interactions. The transition rate of a particle from the 
state $|\bk_i\rangle$ to the state $|\bk_j\rangle$, $T_{ji}$, caused by this interaction should 
be proportional to $a^\dagger_{\bk_j}a_{\bk_i}$. Using (\ref{defaaGS}), we obtain
$T_{ji}\propto(n_{\bk_j}+1)n_{\bk_i}$, like in
the case of bosons, with the extra condition that $T_{ji}=0$ if
$n_{\bk_j}\ge\xi$. These transition rates obey the DBE,
like in Section \ref{BFgas}.

The thermal properties of the GS have been calculated in several
papers (see for example
Ref. \cite{JMathPhys.11.2691.1970.Katsura}). We assume that the
single-particle states $|\bk_i\rangle$ are eigenstates of the
Hamiltonian of the (ideal) system, with the eigenvalues $\epsilon_i$
(or, equivalently, $\epsilon_{\bk_i}$). Since only up to $\xi$
particles can occupy at same time such a state, the partition function can be written as
%
\numparts
\begin{eqnarray} \label{GSfunct}
\cZ^{(GS)} &=& \prod_i\left[\sum_{j=0}^\xi
  e^{j\beta(\mu-\epsilon_i)}\right] \equiv \prod_i\left[
\frac{1-e^{(\xi+1)i\beta(\mu-\epsilon_i)}}{1-e^{\beta(\mu-\epsilon_i)}}
\right] 
\label{cZPSn}
\end{eqnarray}

From Eq. (\ref{cZPSn}) we can calculate the grandcanonical thermodynamic
potential, 
\begin{equation}
\Omega^{(GS)} = -k_BT\log(\cZ^{(GS)})
\end{equation}
and the total particle number,
$N=\partial\log(\cZ^{(GS)})/\partial(\beta\mu)\equiv\sum_i \tilde
n_i^{(GS)}$, where
\begin{equation}
\tilde n_i^{(GS)} = \frac{\sum_{n=0}^\xi ne^{n\beta(\mu-\epsilon_i)}}
{\sum_{n=0}^\xi e^{n\beta(\mu-\epsilon_i)}} =
\frac{1}{e^{\beta(\epsilon_i-\mu)}-1} -
\frac{\xi+1}{e^{(\xi+1)\beta(\epsilon_i-\mu)}-1} 
\label{Ntot}
\end{equation}
\endnumparts
%
is the average population of the single-particle level $i$. From the
Eqs. (\ref{GSfunct}), we can calculate all the thermodynamics of the
system. 

At this point the differences between FES and GS become clear.
We can split the GS system also into species, like the FES system. 
Although in both statistics the maximum number of particles
\textit{per} species is the same 
(provided that the species have the same dimensions), 
the number of microconfigurations is different in general.
This has been calculated in several papers, e.g. 
\cite{NuovoCim.17.495.1940.Gentile,RevBrasFis.6.471.1976.Ponczek,AmJPhys.30.49.1962.Fisher}.
If in the species $i$ of a GS system we have, like in FES, $G_i$ states and $N_i$ particles, 
and we denote by $R_i\equiv[G_i/(\xi+1)]$ the greatest integer contained in 
$G_i/(\xi+1)$, then the number of microconfigurations in which the particles can be 
distributed in the species $i$ is 
\begin{equation}
W_{GS}(G_i,N_i) = \sum_{r=0}^R(-1)^r\frac{G_i[G_i+N_i-(\xi+1)r-1]!}
{r!(G_i-r)![N_i-(\xi+1)r]!}. \label{WGS}
\end{equation}
Expression (\ref{WGS}) is different, in general, from Eq. (\ref{Bose_microC}), 
although it reduces to it in special cases, like e.g. when $\xi\ge G_i$ 
in (\ref{WGS}) and $\alpha=0$ in (\ref{Bose_microC}). 

These differences are reflected in the thermodynamic properties of GS and FES systems,
a fact which is particularly obvious in systems with constant density of states. 

A transition rate between different species in GS would be rather difficult to 
write, since it depends on how many states in the target species are 
fully occupied. Therefore for GS systems we shall consider just transitions 
from one state to another, as discussed in the beginning of this section.

The main thermodynamic quantities of GS have been calculated 
for example in
Ref. \cite{JMathPhys.11.2691.1970.Katsura}. In the next section we will 
recover these results by Monte Carlo simulations.

\section{Numerical implementation}

In order to simulate the time evolution of the system and, in the same time, to 
validate numerically the transition probabilities introduced in 
Section \ref{formalism}, we take as model system the $d$D 
ideal gas, contained in a box of linear dimension $L$, with the density of one-particle states
\begin{equation}
g(\epsilon) = \frac{\hbar^2}{2 m} \left(\frac{L}{2 \pi}\right)^d 
                  \frac{\pi^{d/2}}{\Gamma(d/2)} \epsilon^{d/2-1}.
\end{equation}
We split the energy axis into ``short'' segments, $(\epsilon_i,\epsilon_{i+1})$,
$i=0,1,\ldots$, which define the particle species. The segments are chosen in such a 
way that each species contains the same number of single particle states, 
$G=\int_{\epsilon_i}^{\epsilon_{i+1}}g(\epsilon)d\epsilon$, for any $i$. 
We shall consider only the case $\alpha_{ij} = \alpha \delta_{ij}$.
For the convenience of numerical calculations, we require that $G/\alpha$ is an integer, while for 
GS we take $G=1$.

We set the total number of particles in each FES system to be $N$ and we fix the energy 
scale by setting the Fermi energy in each system. 
For the numerical calculations, we have 
to fix the highest energy in the system, $\epsilon_{\rm max}$, so that the contribution coming 
from the higher energy levels can be neglected in the thermodynamic results.

In this paper we shall consider only 1D and 2D systems, so
we define a scaling temperature,
$T_0$, by $\rho \lambda^d(T_0) = 1$,
where $\rho = N / L^d$ is the $d$D particle density and
$\lambda(T) = h / \sqrt{2 \pi m k_B T}$ is the {\it thermal length}. 
This gives
\begin{equation}
T_0 = \frac{N^{2/d}}{L^2}\left(\frac{h^2}{2 \pi m k_B}\right)^{1/2} ,\ 
d=1,2. \label{scalingT}
\end{equation}

Concretely, we shall take $\epsilon_{\rm F}(\alpha)\equiv\mu(T=0;\alpha)=\alpha$ and $\epsilon_{\rm max}=30$, 
which would permit us to vary the temperature in a range 
between zero and $2T_0$, without any observable effect on the
thermodynamic quantities. We set the energy unit to 
$\epsilon_{\rm F}(\alpha=1)$.

In FES systems we cannot use directly the transition rates given in
Eqs. (\ref{TBji}) and (\ref{TFji}), because they involve in general too big numbers
which cannot be calculated numerically. The quantities that are interesting for us are
actually the relative transition rates, $T_{p,ji}(N_j-1,N_i)/T_{p,ij}(N_i,N_j-1)$. 
Moreover, in order to obtain a common description for both, bosons and fermions,
we define $G_{0i}\equiv G_i+\alpha N_i$, for bosons, and 
$G_{0i}\equiv G_i-(1-\alpha)N_i+1$ for fermions. In these notations 
the number of microconfigurations for the species $i$ becomes 
\cite{PhysRevLett.73.922.1994.Wu,PhysRevB.60.6517.1999.Murthy}
\begin{equation}
W(G_i,N_i) = \frac{[G_{0i}+(1-\alpha) N_i -1]!}{N_i!(G_{0i}-\alpha N_i -1)!}
\label{WBFG0}
\end{equation}
for any type of particles. Therefore from now on we shall omit the subscripts $B$,
$F$, or $p$ in our notations.

To calculate the ratio $T_{p,ji}(N_j-1,N_i)/T_{p,ij}(N_i,N_j-1)$ we take its
logarithm and, by noticing from Eqs. (\ref{TBji}) and (\ref{TFji}) that 
$T_{p,ji}(N_j-1,N_i)=T_{p,ij}(N_i-1,N_j)$, we write 
\begin{eqnarray}
\log\left[\frac{T_{ji}(N_j-1,N_i)}{T_{ij}(N_i,N_j-1)}\right] &=&
\log[T_{ij}(N_i-1,N_j)]-\log[T_{ij}(N_i,N_j-1)] \nonumber \\ 
&\approx& 
-\frac{\partial\log[T_{ij}(N_i,N_j)]}{\partial N_i}
+\frac{\partial\log[T_{ij}(N_i,N_j)]}{\partial N_j}
\nonumber \\
&\approx& \log\left\{\frac{
\frac{[1+(1-\alpha)\tilde{n}_j]^{1-\alpha}(1-\alpha\tilde{n}_j)^\alpha}{\tilde{n}_j}e^{-\beta\epsilon_j}}
{\frac{[1+(1-\alpha)\tilde{n}_i]^{1-\alpha}(1-\alpha\tilde{n}_i)^\alpha}{\tilde{n}_i}e^{-\beta\epsilon_i}}\right\}
\label{relativeT}
\end{eqnarray}
where $\tilde{n}_i\equiv N_i/G_i$ and in obtaining the last line we used the 
Stirling approximation, $\log N!\approx N\log(N/e)$. Therefore, from now on we shall use 
\begin{equation}
T_{ji}(N_j,N_i) = \tilde{n}_i[1+(1-\alpha)\tilde{n}_j]^{1-\alpha}(1-\alpha \tilde{n}_j)^\alpha ,
\label{Tjilog}
\end{equation}
up to the relative canonical factor, $e^{\beta(\epsilon_i-\epsilon_j)}$. 
One can understand the derived transition rates as follows:
a new state is proposed by moving one particle from 
species $i$ to species $j$ with a {\it step} probability which depends
on the instantaneous average occupations of the two species,
i.e. $\sim \tilde{n}_i[1+(1-\alpha)\tilde{n}_j]^{1-\alpha}
(1-\alpha \tilde{n}_j)^\alpha$,  
and it is {\it accepted} with the usual Metropolis probabilities,
$\min(1,\exp(\beta(\epsilon_i-\epsilon_j))$.
In this way the dynamics of FES systems can be viewed as 
the dynamics of classical ideal gas, but with new
{\it step} probabilities which stem from a generalized exclusion principle.   
This picture is consistent with the one described in Ref.\ 
\cite{JStatMech.2009.P02021.2009.Guastella} for Bose and Fermi systems.
From Eq. (\ref{relativeT}) or (\ref{Tjilog}) it becomes obvious that 
when equilibrium is reached, all the transition rates become equal and 
therefore the DBE is satisfied, as shown in Section \ref{FESsection}.

In the numerical calculations we shall take $G_{0i}\equiv G=120$ for any $i$
and $N=12000$.
These values are large enough to 
obtain the analytical equilibrium distribution with high accuracy.
Nevertheless, a discussion about the species' size is also given below.
Using these parameters, the considered systems are proved to be 
very well described analytically by continuous models, 
so that a direct comparison is possible.

\begin{figure}
\begin{center}
\includegraphics[width=9.0cm]{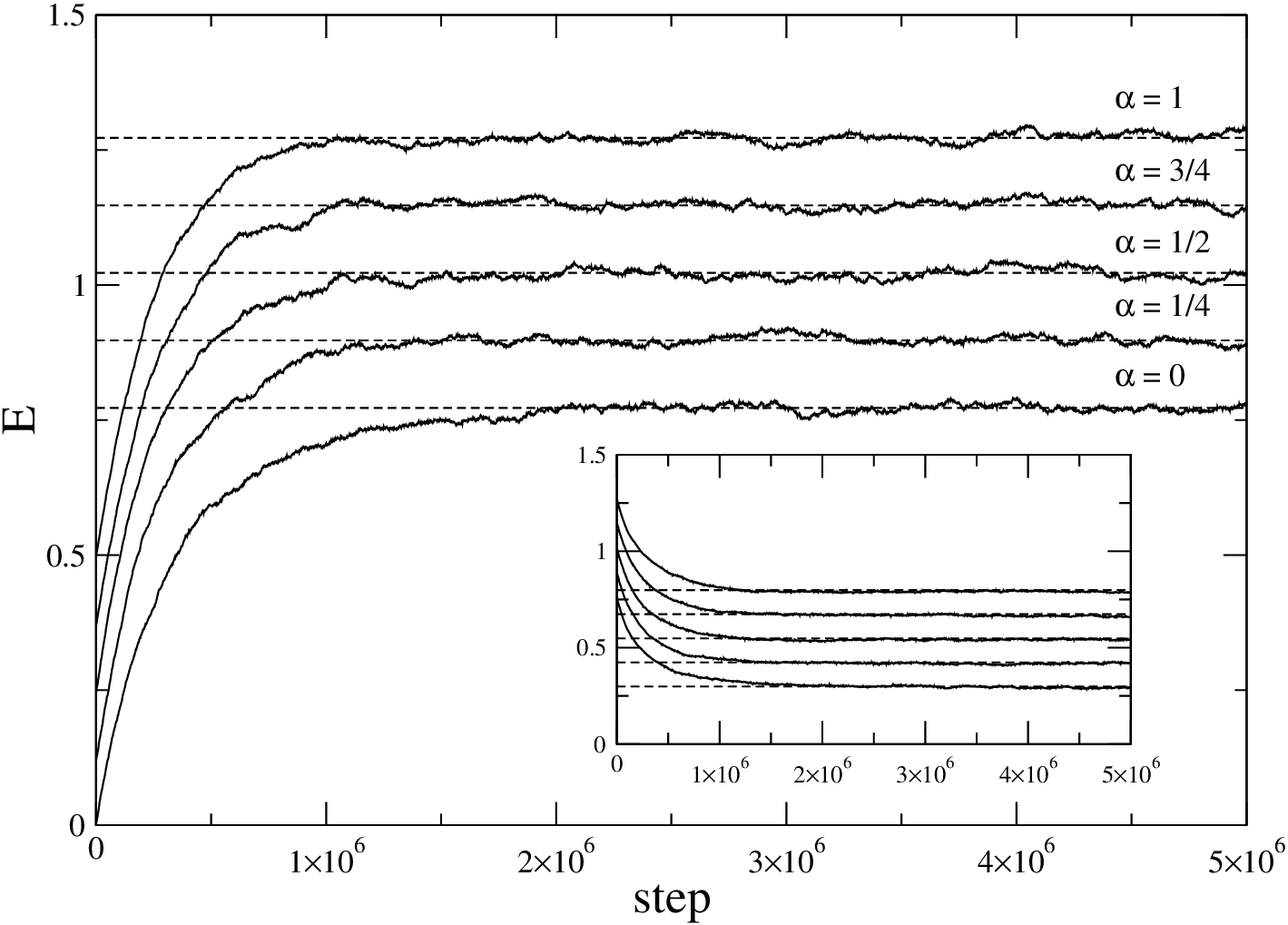}
\end{center}
\caption{Typical equilibration of two-dimensional FES systems,
by raising (main plot) or lowering (inset) the temperature. 
The horizontal lines mark the average equilibrium energy of each system,
calculated analytically.}
\label{equil}
\end{figure}

Typical results for equilibration in two-dimensional systems with the 
FES parameter 
$\alpha$ taking the values 0, 1/4, 1/2, 3/4, 1 are indicated 
in Fig.\ \ref{equil}.
The systems are first prepared in their ground state ($T=0K$) configurations
and the dynamics towards equilibrium is depicted at $T=T_0$. 
The inset contains the opposite situation i.e. the systems are quenched from
$T_0$ to $0.5 T_0$. 
In this case, the initial configuration is
a random configuration drawn according to the equilibrium distribution,
calculated analytically at $T_0$.
In both cases, i.e. either increasing or decreasing the temperature, 
after a number 
of $\sim 10^6$ steps, the energy of systems gets close to and 
starts oscillating around the equilibrium value, as calculated analytically.

\begin{figure}
\begin{center}
\includegraphics[width=9.0cm]{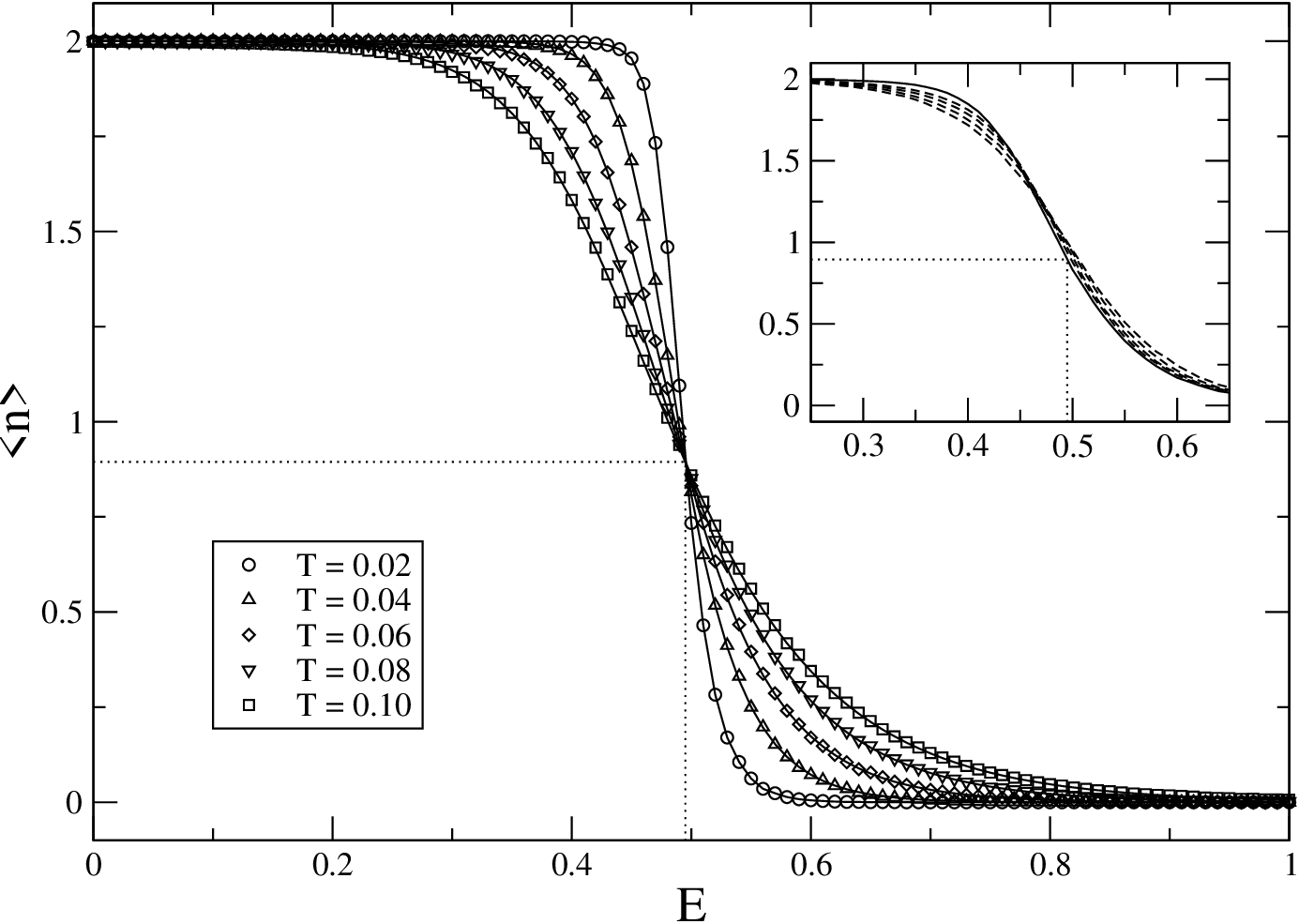}
\end{center}
\caption{Main plot: Analytical (solid lines) and numerical (symbols) data
for a two-dimensional system with a number of single particle states 
in a species $G=120$,
for the five indicated temperatures.
Inset: Numerical data for $T=0.06T_0$, with 
$G=120$ (solid) and
$G=1, 2, 4, 8$ (dashed),
where the largest deviation corresponds to $G=1$.
Here the symbols have been dropped for convenience.
}
\label{species}
\end{figure}

\begin{figure}
\begin{center}
\includegraphics[width=9.0cm]{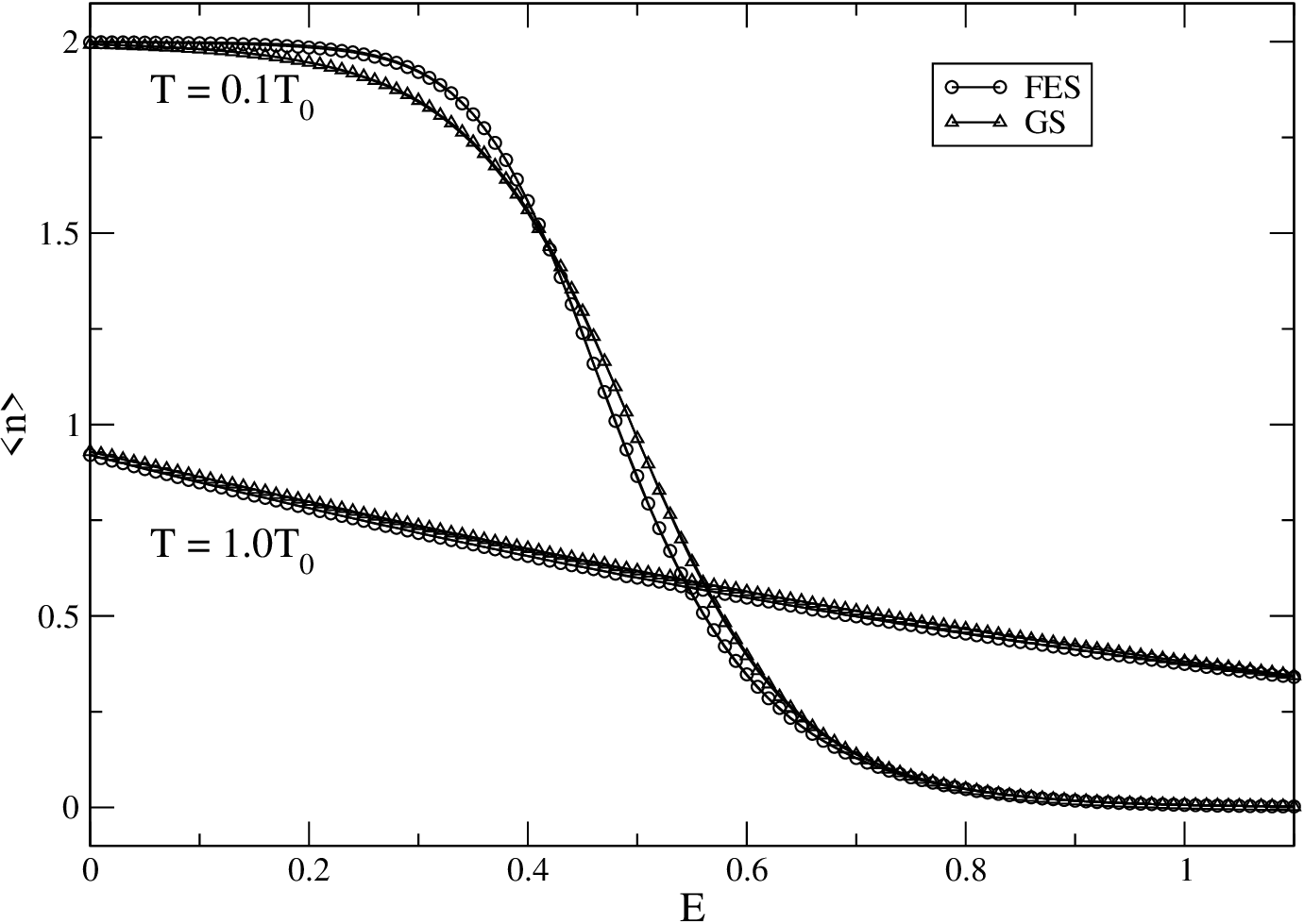}
\end{center}
\caption{Comparison between FES and GS for semions 
($\alpha = 1/\xi = 1/2$),
for low ($T=0.1T_0$) and high temperatures ($T=T_0$). Analytical data is 
represented by solid lines and numerical data by symbols.
}
\label{FES_GS}
\end{figure}
 
\begin{figure}[t]
\begin{center}
\includegraphics[width=7.0cm]{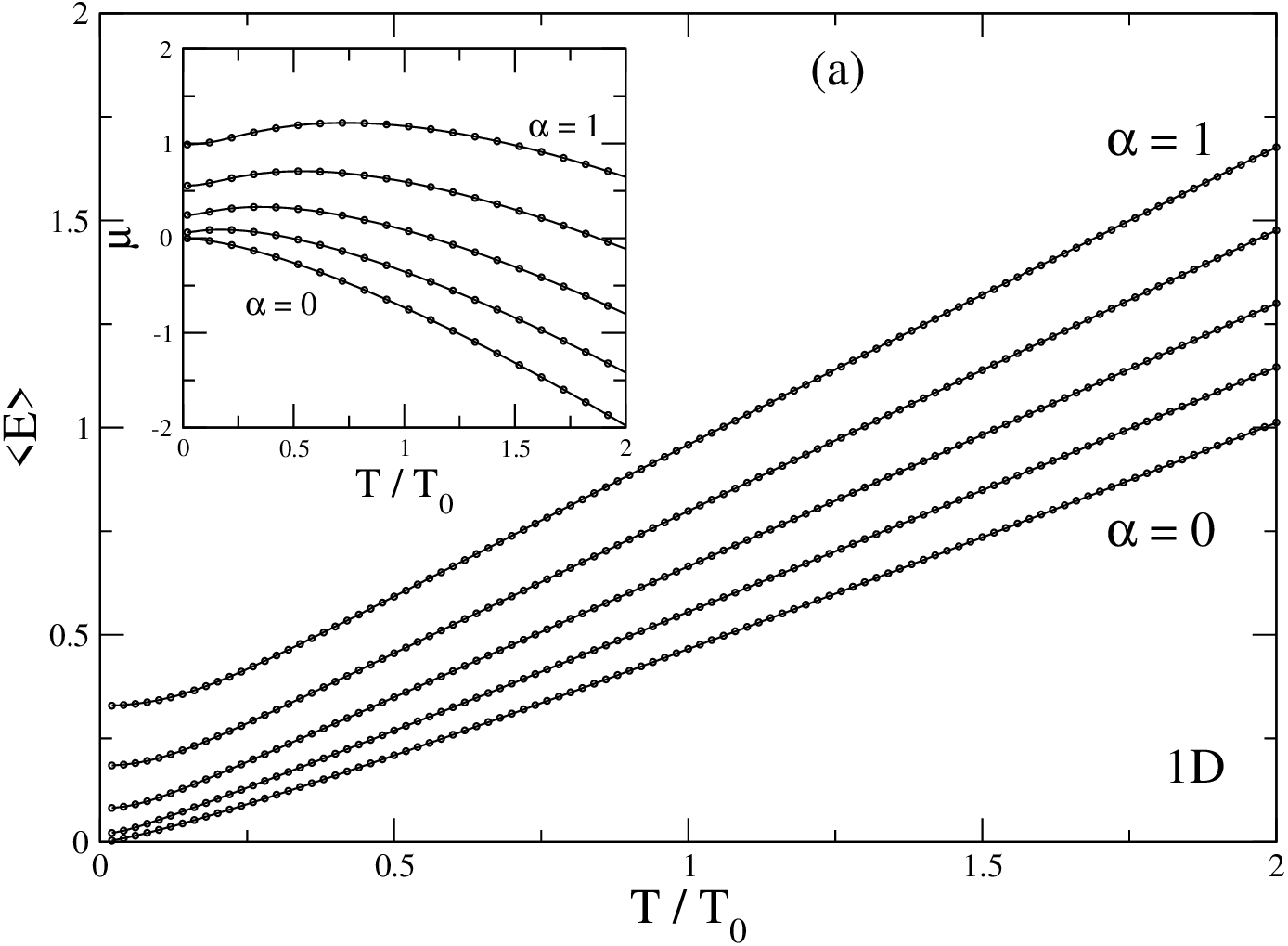} 
\includegraphics[width=7.0cm]{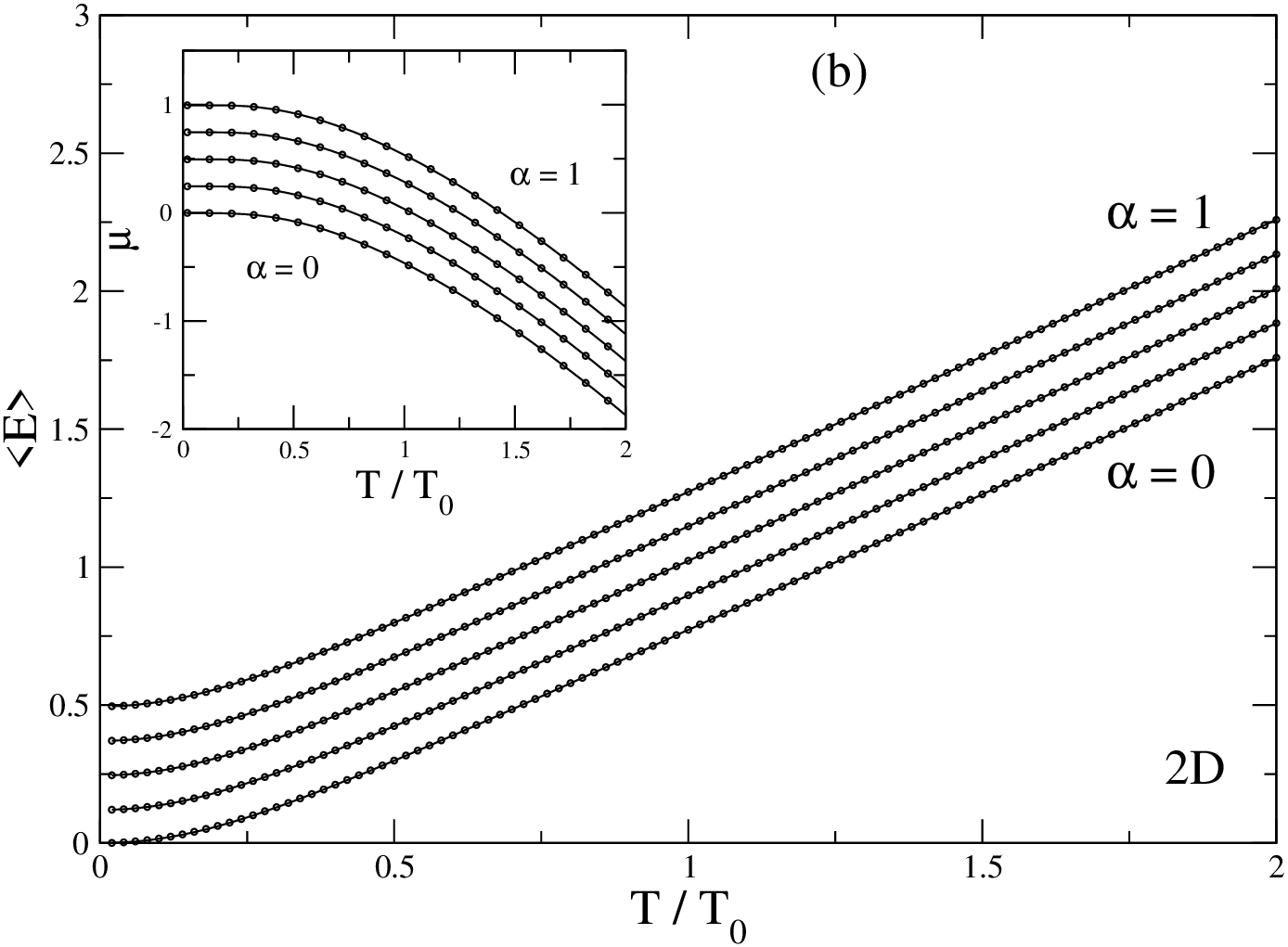}
\end{center}
\caption{Average total energy vs. temperature (symbols) for $d=1$ (a) and 
$d=2$ (b) for FES systems with $\alpha=0, 1/4, 1/2, 3/4, 1$. 
Inset: Temperature dependence of the chemical potential. 
Solid lines represent analytical calculations using 
the corresponding continuous model.
}
\label{E_mu}
\end{figure}

\begin{figure}[t]
\begin{center}
\includegraphics[width=7.0cm]{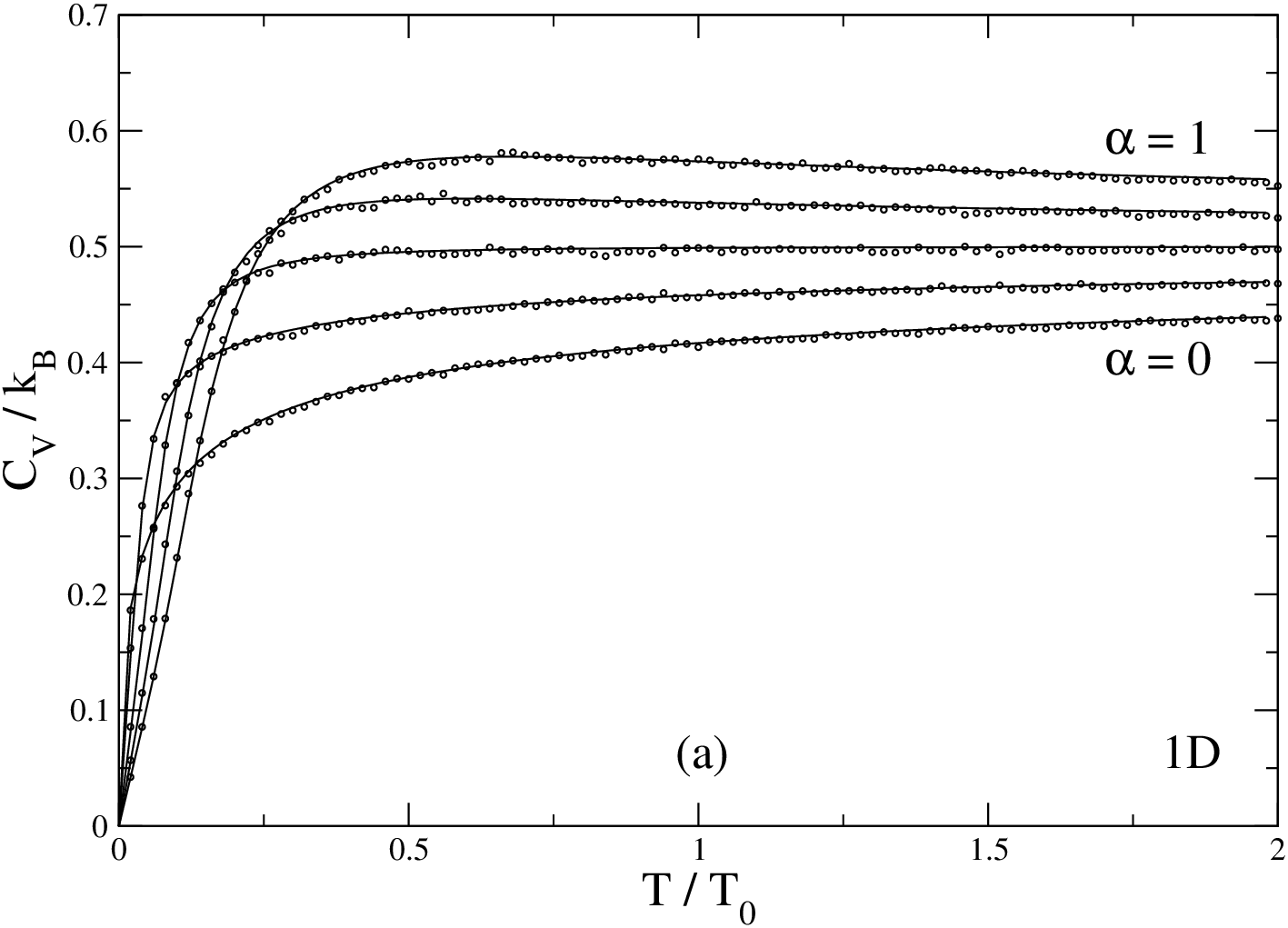} 
\includegraphics[width=7.0cm]{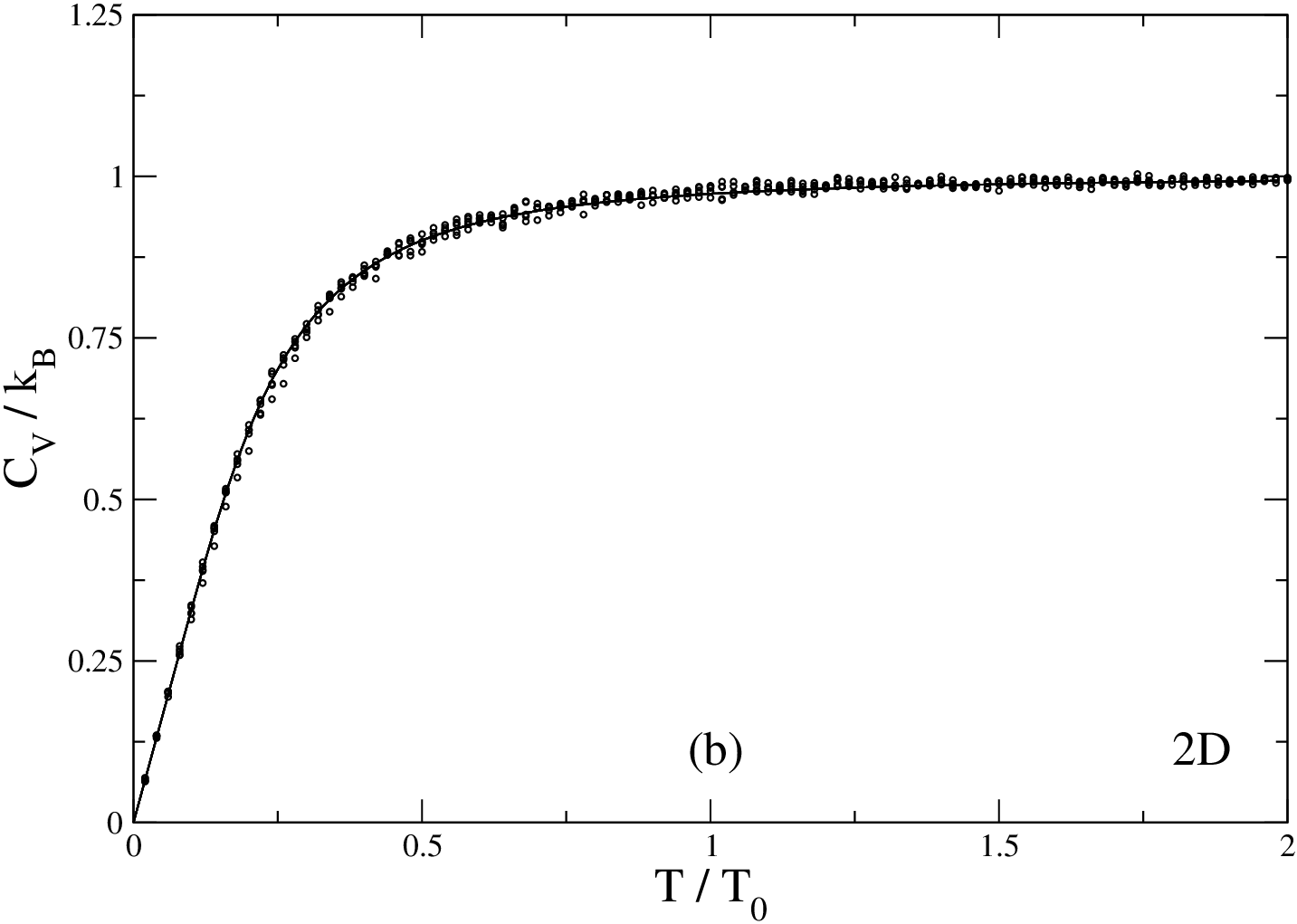}
\includegraphics[width=7.0cm]{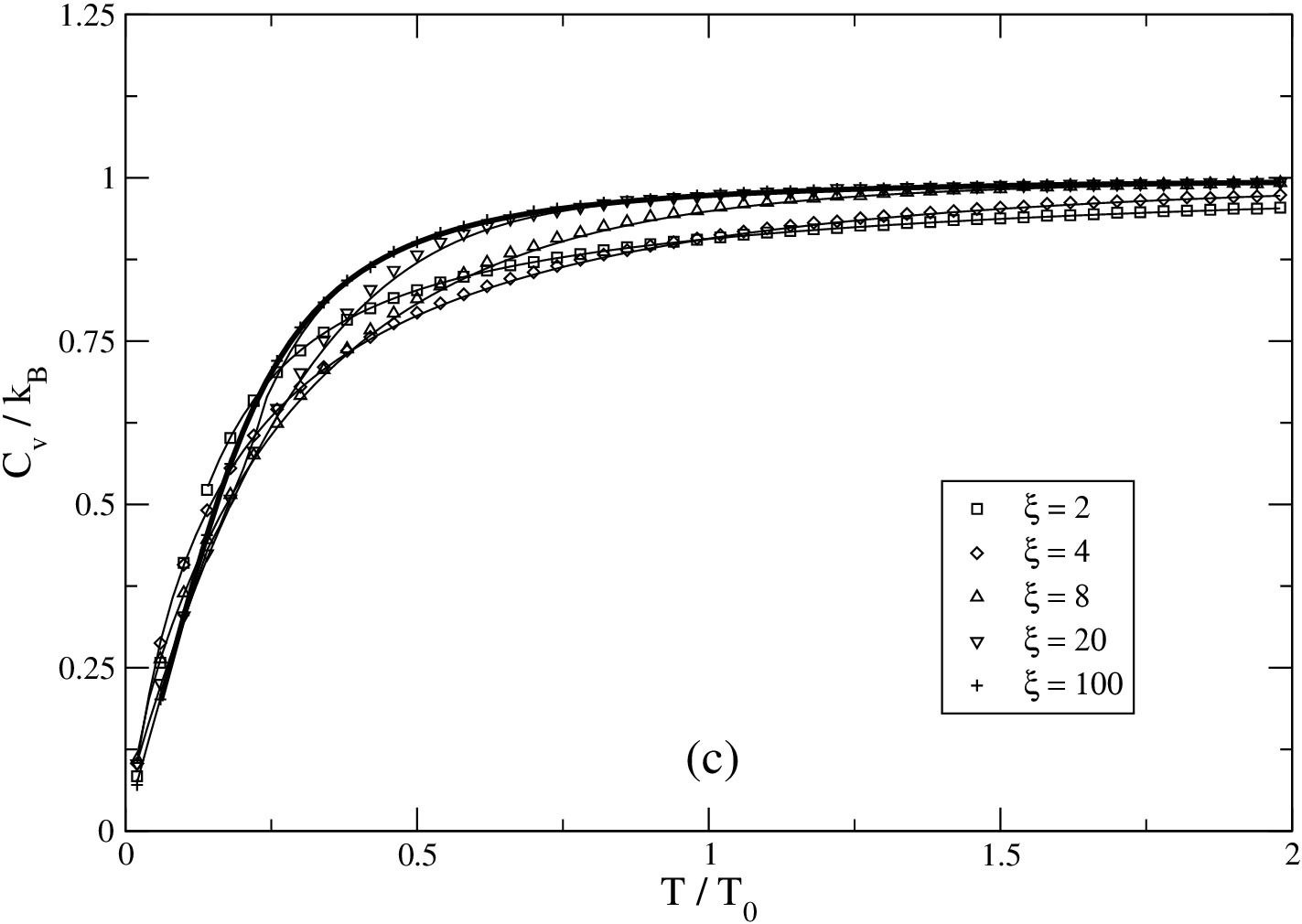}
\end{center}
\caption{Heat capacity vs. temperature (symbols) for one-dimensional (a) and two-dimensional (b) FES systems, and for two-dimensional GS systems (c). Solid lines represent analytical calculations using the corresponding continuous model.
}
\label{Cv}
\end{figure}

Figure \ref{species} shows the average occupation of a two-dimensional 
system with $\alpha=1/2$, which is calculated 
performing time averages
after the system has reached equilibrium. The system is first very well 
thermalized for number of $N_0=10^9$ steps and 
then the time average is obtained using an equal number of additional steps. 
We included every 1000$^{\rm th}$ microconfiguration in the time-average,
i.e. we considered a total of $10^6$ microconfigurations.

The data is fitted very well by analytical calculations using 
the corresponding continuous model.  
From Eq. (17) of Ref. \cite{PhysRevLett.73.922.1994.Wu},
one can see that in the degenerate limit, 
i.e. low temperatures,
the different distributions cross at the point 
($E = \mu_{T=0}=0.495,\langle n \rangle = \sqrt{4/5}$), 
as it is indicated by dotted lines, the small deviation of 0.005 in 
$\mu_{\mbox{\scriptsize T=0}}$, 
being caused by the finite spacing between the energy levels.
In FES systems, if one decreases the number of 
single particle states in a species, $G$, the average 
distributions deviate, as it is shown in the inset.
For the limiting case, $G=1$,
 it is established that 
the average FES distributions for different temperatures 
in the degenerate limit cross at $\langle n \rangle = 1$.
Therefore we emphasize the high importance of the species' size 
in obtaining the correct FES distribution. 

In Fig. \ref{FES_GS} we analyze comparatively the two statistics, FES and GS. 
For this we plot the equilibrium distributions for semions ($\alpha = 1/\xi = 1/2$).
Although in the high temperature limit
the distributions become both identical with the classical 
Maxwell Boltzmann distribution, 
there are visible differences at low temperatures. 
One particular difference is that GS equilibrium distributions 
in the low temperature limit cross at $\langle n \rangle = \xi/2$,
which is not the case for the FES systems.

Next we perform long runs to obtain thermodynamic quantities 
with good accuracy. 
As before, we take $N_0=10^9$ steps to reach equilibrium and another 
$N_1=10^{10}$ steps to obtain the time averages. The results for the 
average energy and chemical potential as function of temperature 
are depicted in Figs. \ref{E_mu}.
The chemical potential is extracted as a fit parameter for the average 
distribution obtained numerically, in order to match the analytical result
for the mean occupation.
The characteristics related to the dimensionality of the systems are 
clearly visible: for $d=2$ case the data is only vertically shifted,
which is a consequence of the thermodynamic equivalence of systems of 
constant density of states and any statistics \cite{ProcCambrPhilos42.272.1946.Auluc,PhysRev.135.A1515.1964.May,PhysRevE.55.1518.1997.Lee,JPA35.7255.2002.Anghel}.

The heat capacity is calculated from the fluctuations of the total energy,
$C_{\mbox{\scriptsize v}} =  
(\langle E^2 \rangle - \langle E \rangle^2) / 
(k_{\mbox{\scriptsize B}} T^2)$. 
The data plotted in Fig.\ \ref{Cv} confirms again the correctness 
of our approach for the proposed step probabilities for FES 
and GS systems. 
As expected, for FES systems and $d=2$, the data for all values 
$\alpha$ collapses on the same curve. 
By contrast, the thermodynamic behavior
of GS systems is no longer equivalent with respect to the $\xi$ parameter.
The thermodynamic equivalence of the 2D systems suggests that FES is a more
natural extension of Fermi and Bose statistics.

\section{Conclusions}

The present approach introduces a stochastic model for the non-equilibrium 
dynamics of systems which obey generalized statistics, 
such as fractional exclusion statistics (FES) and 
Gentile statistics (GS).
The model has at its core a time-dependent Markov chain
in the microconfiguration state space,
which generalizes previous results from Ref. \cite{JStatMech.2009.P02021.2009.Guastella}.
In the derivation of the stochastic 
transition probabilities, a division can be made which separates them
into {\it acceptance} probabilities, corresponding to
transition probabilities in a classical ideal gas 
(i.e. Metropolis probabilities), 
and {\it step} probabilities, which account for
the generalized exclusion principle employed.  

The obtained probabilities are tested extensively 
by Monte Carlo simulations on one- and 
two-dimensional FES and GS systems. The numerical results reflecting 
several thermodynamic quantities overlap 
very well with the reference data, calculated analytically.
We point out here the crucial role played by the size of individual species
in the FES systems, as well as the fundamental differences which appear 
in the thermodynamics properties of the two-dimensional FES and GS systems.
 
The Monte Carlo approach to FES could yield new insights 
in the time evolution of {\it finite} quantum many-body systems 
regarded as FES systems,
indicating potential differences related to out-of-the-equilibrium 
phenomena.


\section*{References}

\begin{thebibliography}{10}

\bibitem{NuovoCim.17.495.1940.Gentile}
G.~Gentile.
\newblock {\em Nuovo Cim.}, 17:493, 1940.

\bibitem{JMathPhys.11.2691.1970.Katsura}
S.~Katsura, K.~Kaminishi, and S.~Inawashiro.
\newblock {\em J. Math. Phys.}, 11:2691, 1970.

\bibitem{RevBrasFis.6.471.1976.Ponczek}
R.~L. Ponczek and C.~C. Yan.
\newblock {\em Revista Brasileira de Fisica}, 6:471, 1976.

\bibitem{PhysRevLett.67.937.1991.Haldane}
F.~D.~M. Haldane.
\newblock {\em Phys. Rev. Lett.}, 67:937, 1991.

\bibitem{JStatMech.2009.P02021.2009.Guastella}
I.~Guastella, L.~Bellomonte, and R.~M. Sperandeo-Mineo.
\newblock {\em J. Stat. Mech.}, 2009:P02021, 2009.

\bibitem{EPL.87.60009.2009.Anghel}
D.~V. Anghel.
\newblock {\em EPL}, 87:60009, 2009.
\newblock arXiv:0906.4836.

\bibitem{PhysRevLett.73.922.1994.Wu}
Yong-Shi Wu.
\newblock {\em Phys. Rev. Lett.}, 73:922, 1994.

\bibitem{PhysRevLett.73.2150.1994.Isakov}
S.~B. Isakov.
\newblock {\em Phys. Rev. Lett.}, 73(16):2150, 1994.

\bibitem{JPhysA.40.F1013.2007.Anghel}
D.~V. Anghel.
\newblock {\em J. Phys. A: Math. Theor.}, 40:F1013, 2007.
\newblock arXiv:0710.0724.

\bibitem{EPL.90.10006.2010.Anghel}
D.~V. Anghel.
\newblock {\em EPL}, 90:10006, 2010.
\newblock arXiv:0909.0030.

\bibitem{ProcCambrPhilos42.272.1946.Auluc}
F.~C. Auluck and D.~S. Kothari.
\newblock {\em Proc. Cambridge Philos. Soc.}, 42:272, 1946.

\bibitem{PhysRev.135.A1515.1964.May}
Robert~M. May.
\newblock {\em Phys. Rev.}, 135:A1515, 1964.

\bibitem{PhysRevE.55.1518.1997.Lee}
M.~Howard Lee.
\newblock {\em Phys. Rev. E}, 55:1518, 1997.

\bibitem{PhysA304.421.2002.Lee}
M.~H. Lee and J.~Kim.
\newblock {\em Physica A}, 304:421, 2002.

\bibitem{JPA35.7255.2002.Anghel}
D.~V. Anghel.
\newblock {\em J. Phys. A: Math. Gen.}, 35:7255, 2002.

\bibitem{JPA39.4787.2006.Anghel}
D.~V. Anghel.
\newblock {\em J. Phys. A: Math. Gen.}, 39:4787, 2006.

\bibitem{PhysLettA.372.5745.2008.Anghel}
D.~V. Anghel.
\newblock {\em Phys. Lett. A}, 372:5745, 2008.
\newblock arXiv:0710.0728.

\bibitem{RJP.54.281.2009.Anghel}
D.~V. Anghel.
\newblock {\em Rom. J. Phys.}, 54:281, 2009.
\newblock arXiv:0804.1474.

\bibitem{NewDevIntSys.1995.Bernard}
D.~Bernard and Y.~S. Wu.
\newblock In M.~L. Ge and Y.~S. Wu, editors, {\em New Developments on
  Integrable Systems and Long-Ranged Interaction Models}, page~10. World
  Scientific, Singapore, 1995.
\newblock cond-mat/9404025.

\bibitem{PhysRevB.56.4422.1997.Sutherland}
B.~Sutherland.
\newblock {\em Phys. Rev. B}, 56:4422, 1997.

\bibitem{PhysRevLett.73.3331.1994.Murthy}
M.~V.~N. Murthy and R.~Shankar.
\newblock {\em Phys. Rev. Lett.}, 73:3331, 1994.

\bibitem{PhysRevLett.74.3912.1995.Sen}
D.~Sen and R.~K. Bhaduri.
\newblock {\em Phys. Rev. Lett.}, 74:3912, 1995.

\bibitem{RomRepPhys59.235.2007.Anghel}
D.~V. Anghel.
\newblock {\em Rom. Rep. Phys.}, 59:235, 2007.
\newblock cond-mat/0703729.

\bibitem{AmJPhys.30.49.1962.Fisher}
M.~E. Fisher.
\newblock {\em Am. J. Phys.}, 29:49, 1961.

\bibitem{PhysRevB.60.6517.1999.Murthy}
M.~V.~N. Murthy and R.~Shankar.
\newblock {\em Phys. Rev. B}, 60:6517, 1999.

\end{thebibliography}
\end{document}